\documentclass[12pt]{article}
\usepackage{graphicx}
\usepackage[fleqn]{amsmath}
\usepackage{amsfonts,amssymb,cite}
\usepackage{pifont}
\usepackage{bm}
\usepackage{subfigure}
\usepackage{enumerate}
\usepackage{epsfig}
\usepackage{soul}
\usepackage{verbatim,setspace}
\usepackage{tikz}
\usetikzlibrary{calc}
\usepackage{array,multirow,wrapfig}
\usepackage{makecell}
\usepackage{lineno}
\usepackage{floatrow}

\def\L{\mathcal{L}}
\newcommand{\ttbb}{t\bar{t}b\bar{b}}

\newcommand\pubnumber{based on ArXiv: 1807.02130}
\newcommand\pubdate{\today}

\def\institute{Inter-University Institute for High Energies (IIHE)\\
Vrije Universiteit Brussel, Pleinlaan 2, B-1050 Brussels, Belgium}
\def\support{\footnote{Work supported by Fonds voor Wetenschappelijk Onderzoek Vlaanderen (FWO). This work is presented on behalf of all the authors: J. D'Hondt, A. Mariotti, K. Mimasu, S. Moortgat and C. Zhang.}}

\def\Title#1{\begin{center} {\Large #1 } \end{center}}
\def\Author#1{\begin{center}{ \sc #1} \end{center}}
\def\Address#1{\begin{center}{ \it #1} \end{center}}

\newcommand\pubblock{\rightline{\begin{tabular}{l} \pubnumber\\
         \pubdate  \end{tabular}}}
\newenvironment{Abstract}{\begin{quotation}  }{\end{quotation}}
\newenvironment{Presented}{\begin{quotation} \begin{center} 
             POSTER PRESENTED AT\end{center}\bigskip 
      \begin{center}\begin{large}}{\end{large}\end{center} \end{quotation}}
\def\Acknowledgements{\bigskip  \bigskip \begin{center} \begin{large}
             \bf ACKNOWLEDGEMENTS \end{large}\end{center}}

\def\beq{\begin{equation}}
\def\eeq#1{\label{#1}\end{equation}}
\def\eeqn{\end{equation}}

\def\beqa{\begin{eqnarray}}
\def\eeqa#1{\label{#1}\end{eqnarray}}
\def\eeqan{\end{eqnarray}}

\let\bar=\overbar

\def\L{{\cal L}}

\def\Dslash{\not{\hbox{\kern-4pt $D$}}}
\def\dslash{\not{\hbox{\kern-2pt $\del$}}}

\def\msb{{\bar{\ssstyle M \kern -1pt S}}}

\begin{document}
\begin{titlepage}
\pubblock

\vfill
\Title{Learning to pinpoint effective operators at the LHC: \\[0.7ex]
a study of the $\ttbb$ signature}
\vfill
\Author{ Seth Moortgat\support}
\Address{\institute}
\vfill
\begin{Abstract}
In the context of the 
Standard Model effective field theory (SMEFT),
we study the LHC sensitivity to four fermion operators involving heavy quarks by employing cross section measurements in the $\ttbb$ final state. Starting 
from the measurement of total rates, we progressively exploit kinematical 
information and machine learning techniques to optimize the projected 
sensitivity at the end of Run III. Indeed, in final states with high 
multiplicity containing inter-correlated kinematical information, multi-variate 
methods provide a robust way of isolating the regions of phase space where the 
SMEFT contribution is enhanced. We also show that training for multiple output 
classes allows for the discrimination between operators mediating the 
production of tops in different helicity states. Our projected sensitivities 
not only constrain a host of new directions in the SMEFT parameter space but 
also improve on existing limits demonstrating that, on one hand, $\ttbb$ 
production is an indispensable component in a future global fit for top quark 
interactions in the SMEFT, and on the other, multi-class machine learning 
algorithms can be a valuable tool for interpreting LHC data in this framework.
\end{Abstract}
\vfill
\begin{Presented}
$11^\mathrm{th}$ International Workshop on Top Quark Physics\\
Bad Neuenahr, Germany, September 16--21, 2018
\end{Presented}
\vfill
\end{titlepage}
\def\thefootnote{\fnsymbol{footnote}}
\setcounter{footnote}{0}

\section{Introduction}
The lack of evidence for signatures of new physics at the Large Hadron Collider (LHC) has led to an increased interest in the Standard Model Effective Field Theory (SMEFT) 
as a model-independent approach to interpret experimental measurements in the context of physics Beyond the Standard Model (BSM). Machine learning classifiers are well suited to the task of discriminating between Standard Model (SM) and SMEFT effects, particularly with increasing final state multiplicity and the intrinsically large parameter space of the SMEFT in which many operators can contribute to a given final state. One of the aims of our study is to quantify the potential of these methods not only to optimally constrain SMEFT operators but also to distinguish operators amongst themselves in order to more accurately pinpoint the origin of an observed deviation in the parameter space. We focus our investigation on SMEFT operators that contribute to top pair production in association with two $b$-jets. We will present the possible reach of the $\ttbb$ process at 13 TeV centre-of-mass energy to a set of four-heavy-quark EFT operators of dimension six.

\section{$\boldsymbol{\ttbb}$ in the SMEFT
\label{sec:EFT}}

In the construction of an EFT one extends the SM Lagrangian with operators of dimension larger than four \cite{Leung:1984ni,Buchmuller:1985jz}. Since dimension five operators only generate baryon or lepton number violating couplings, the first extension happens with the addition of dimension six effective operators that are suppressed by the square of an energy scale $\Lambda$: $ \L = \L_{SM} + \sum_{i} \frac{C_{i}}{\Lambda^{2}} \ O_{i} $, where $C_{i}$ is the Wilson coefficient corresponding to the EFT operator $O_{i}$. In this work we absorb the factor $\Lambda^{-2}$ in the definition of the Wilson coefficient.


The most important featureof the $\ttbb$ final state, is its capability of exploring new contact interactions among the third-generation quarks.
There are 10 independent 4-fermion operators involving only heavy quarks that contain $\ttbb$ interactions~\cite{Grzadkowski:2010es}. Following the basis choice recommended by the LHC Top Working Group~\cite{AguilarSaavedra:2018nen}:
\begin{align}
\nonumber
 & O^{1}_{QQ} = \frac{1}{2}\left( \bar{Q} \ \gamma_{\mu} \  Q \right)  \left( \bar{Q}  \ \gamma^{\mu} \  Q \right), \, \, &O^{8}_{QQ} = \frac{1}{2}\left( \bar{Q} \ \gamma_{\mu} \  T^{A} \  Q \right)  \left( \bar{Q}  \ \gamma^{\mu} \  T^{A} \  Q \right), \\
\nonumber  
& O^{1}_{tb} = \left( \bar{t}  \ \gamma_{\mu}  \ t \right)  \left(\bar{b} \ \gamma_{\mu} \  b \right), \, \,  &O^{8}_{tb} = \left( \bar{t}  \ \gamma_{\mu} T^{A} \  \ t \right)  \left(\bar{b} \ \gamma_{\mu} \ T^{A} \  b \right), \\
\nonumber  
& O^{1}_{Qt} = \left( \bar{Q} \ \gamma_{\mu} \  Q \right)  \left( \bar{t}  \ \gamma^{\mu} \  t \right),  \, \,  & O^{8}_{Qt} = \left( \bar{Q} \ \gamma_{\mu} \  T^{A} \  Q \right)  \left( \bar{t}  \ \gamma^{\mu} \  T^{A} \  t \right),  \\
\nonumber
& O^{1}_{Qb} = \left( \bar{Q} \ \gamma_{\mu} \  Q \right)  \left( \bar{b}  \ \gamma^{\mu} \ b \right), \,\, & O^{8}_{Qb} = \left( \bar{Q} \ \gamma_{\mu} \  T^{A} \  Q \right)  \left( \bar{b}  \ \gamma^{\mu} \  T^{A} \  b \right),  \\
\nonumber
& O^{1}_{QtQb} = \left( \bar{Q} \ t \right) \varepsilon \left( \bar{Q} \ b \right) ,  \, \, & O^{8}_{QtQb} = \left( \bar{Q} \  T^{A} \  t \right) \varepsilon \left( \bar{Q} \  T^{A} \  b \right).
 \end{align}

$Q$ represents the left-handed SU(2) doublet of third generation quarks (top and bottom), $t$ and $b$ represent the right-handed top and bottom quarks, $T^{A}$ denotes the $SU(3)$ generators and $\varepsilon$ is the totally antisymmetric Levi-Civita tensor in $SU(2)$-space. Of the first two operators $O^{1}_{QQ}$ and $O^{8}_{QQ}$, only one linear combination can be probed by the four-top process \cite{Zhang:2017mls}. In contrast, in $\ttbb$ production both degrees of freedom are probed independently.
The advantages of an EFT interpretation of $\ttbb$ measurements at the LHC can be summarized as follows:
\begin{itemize}
    \item A sufficiently large inclusive cross section ($\sim 3$ pb) that allows for the use of differential information after 300 fb$^{-1}$ of integrated luminosity.
    \item It directly constrains 6 four heavy quark operators for the first time.
    \item It breaks the degeneracy in a blind direction of the parameter space with respect to four top measurements.
\end{itemize}

\section{Increasing the sensitivity to individual operators}
\label{sec:sensitivity}

\begin{wrapfigure}{l}{.39 \textwidth}
\includegraphics[width=.95\textwidth]{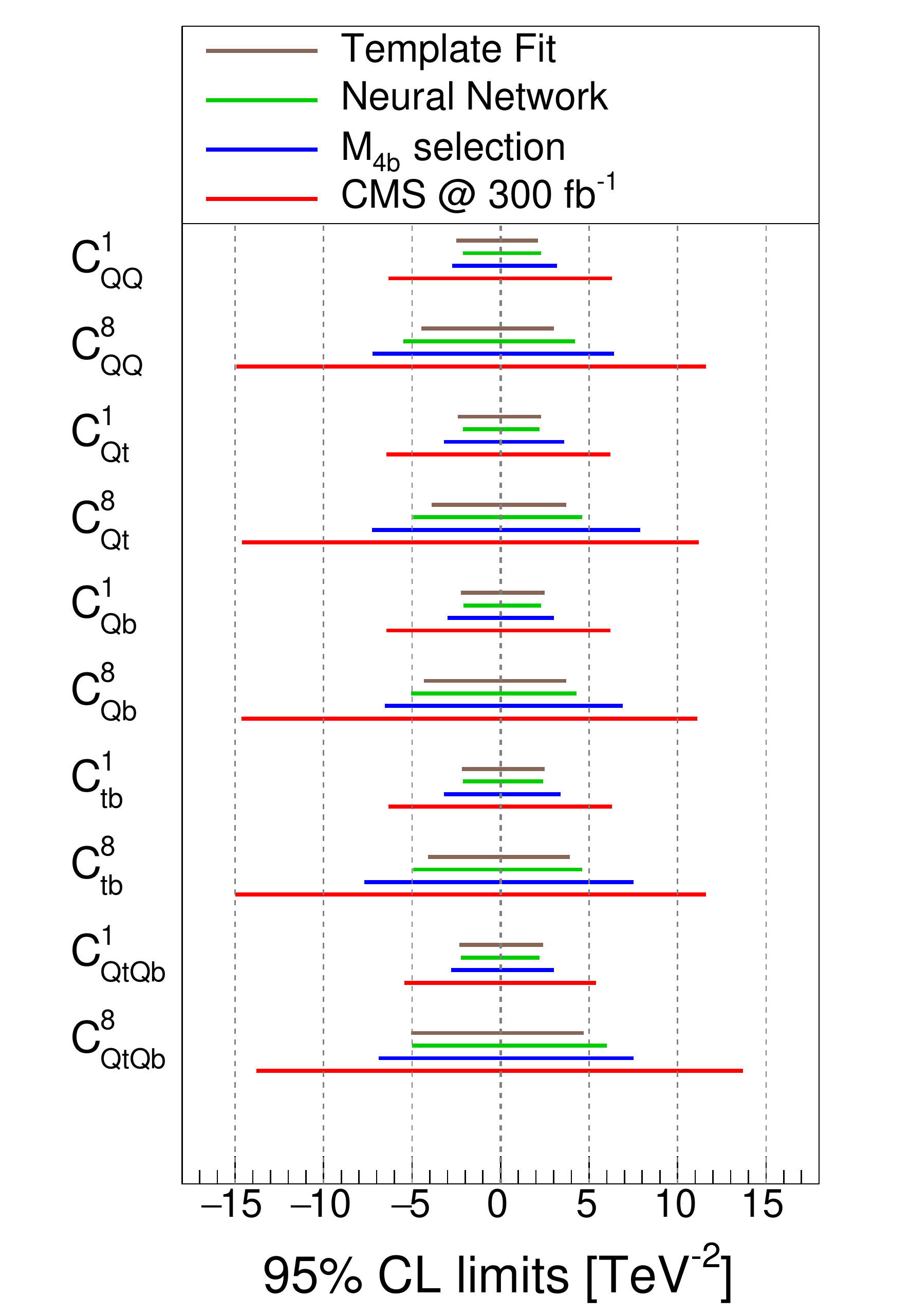}
\caption{\label{fig:summary_Limits}
Summary of the individual limits at 95\% CL on all the Wilson coefficients.
An upper cut on every energy scale of the process of $M_{cut}=2$ TeV has been applied throughout.
}
\end{wrapfigure}

Before highlighting the main novelty of this work in Section \ref{sec:MultipleOperators}, we first perform a sensitivity study in which we turn on one operator at a time and try to obtain an increased sensitivity to the EFT operators. As a starting point, the $\ttbb$ cross section measurement by CMS~\cite{Sirunyan:2017snr} (13 TeV, 2.3 fb$^{-1}$), resulting in a value of $\sigma_{\ttbb,CMS} = 88 \pm 12 \ (stat.) \pm 29 \ (syst.)$ fb, can be translated into individual limits on the Wilson coefficients. We make projections for 300 fb$^{-1}$ of integrated luminosity, by scaling the statistical uncertainty and assuming an overall systematic uncertainty of 10\%. The 95\% CL limits on the Wilson coefficients are shown in red in Fig.~\ref{fig:summary_Limits}. In order to further increase the sensitivity, one needs to employ specific event selections in the reconstructed phase space. We show an improved sensitivity by a  factor of $\sim 2$ can be achieved by making a selection on the invariant mass of the four $b$-jets in the event ($M_{4b}$)~\cite{Degrande:2010kt}, as shown with the blue lines in Fig.~\ref{fig:summary_Limits}. By combining all kinematic information into a neural network (NN) and making a selection on its output, the sensitivity can be even further increased (green lines).
\\
\\

\section{Learning the effective operators
\label{sec:MultipleOperators}}
\begin{figure}[ht!]
\floatbox[{\capbeside\thisfloatsetup{capbesideposition={left,top},capbesidewidth=.4\textwidth}}]{figure}[\FBwidth]
{\caption{\label{fig:2DplotDiscriminators}
 Normalized distributions of the combined NN outputs for events corresponding to SM (red), SM+EFT ($t_L$ operators) (green) and SM+EFT ($t_R$ operators) (blue). The Wilson coefficients are set to 20 (TeV$^{-2}$). The size of each box is proportional to the abundance of events of the corresponding sample.
The discriminators on the x and y axis are as defined in the text.  
}}
{\includegraphics[width=.6\textwidth]{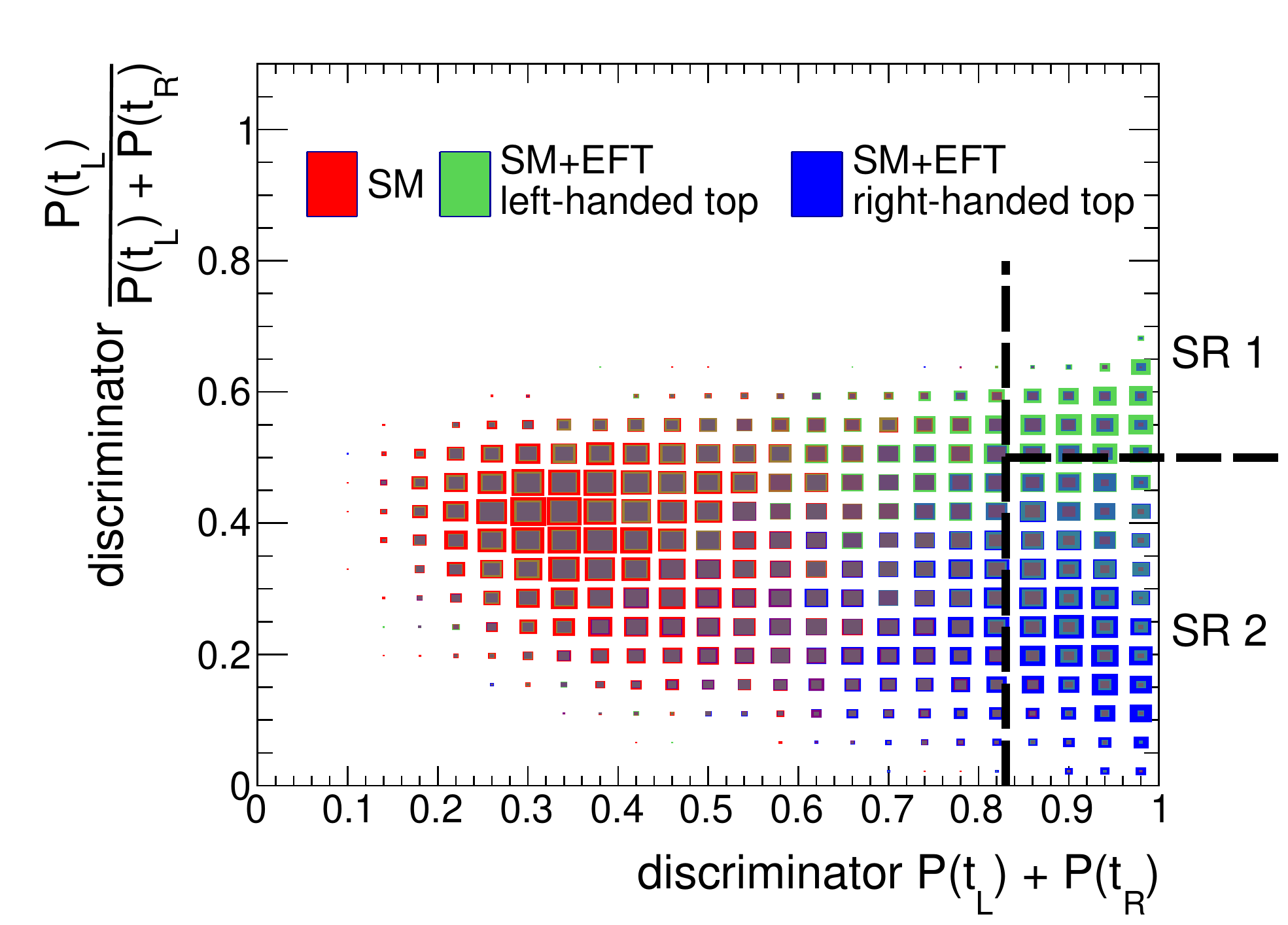}}
\end{figure}

We finally illustrate the strength of a multi-class output structure of the NN, which will allow for a separation between operators with different helicity structures of the top quark currents. We illustrate this with an example using events generated with both $C_{Qb}^{1}$ and $C_{tb}^{1}$ non-zero. We defined a set of 18 kinematical variables, consisting of transverse momenta, invariant masses and angular separations in $\Delta R$ between final-state particles. These variables are fed as input to a shallow neural network with one hidden layer, containing 50 neurons and 3 output classes. The outputs represent the probabilities (P) of an event belonging to one of the following three categories: a Standard Model event (SM), an event from an EFT operator with a left-handed top quark ($t_{L}$) current and an event from an EFT operator with a right-handed top quark ($t_{R}$) current.
To visualize the separation potential of the neural network between the three classes, Fig.~\ref{fig:2DplotDiscriminators} shows how the different classes are distributed in the plane of the combined neural network outputs. The x-axis represents the summed probability $P(t_{L}) + P(t_{R})$ that is able to separate the SM events (red) from any kind of event that includes the insertion of an EFT operator. On the y-axis, the normalized probability \( \frac{P(t_{L})}{P(t_{L}) + P(t_{R})} \) is displayed, designed to distinguish between the $t_{L}$ (green) and the $t_{R}$ (blue) categories. These distributions show a clear concentration of SM events to the left, whereas the $t_{L}$ and $t_R$ contributions dominantly populate the upper and lower right hand corners, respectively.
 We therefore define two signal regions (SR1) and (SR2) as delimited in Fig.~\ref{fig:2DplotDiscriminators}.


\begin{figure}[ht!]
\center
\includegraphics[width=.4\textwidth]{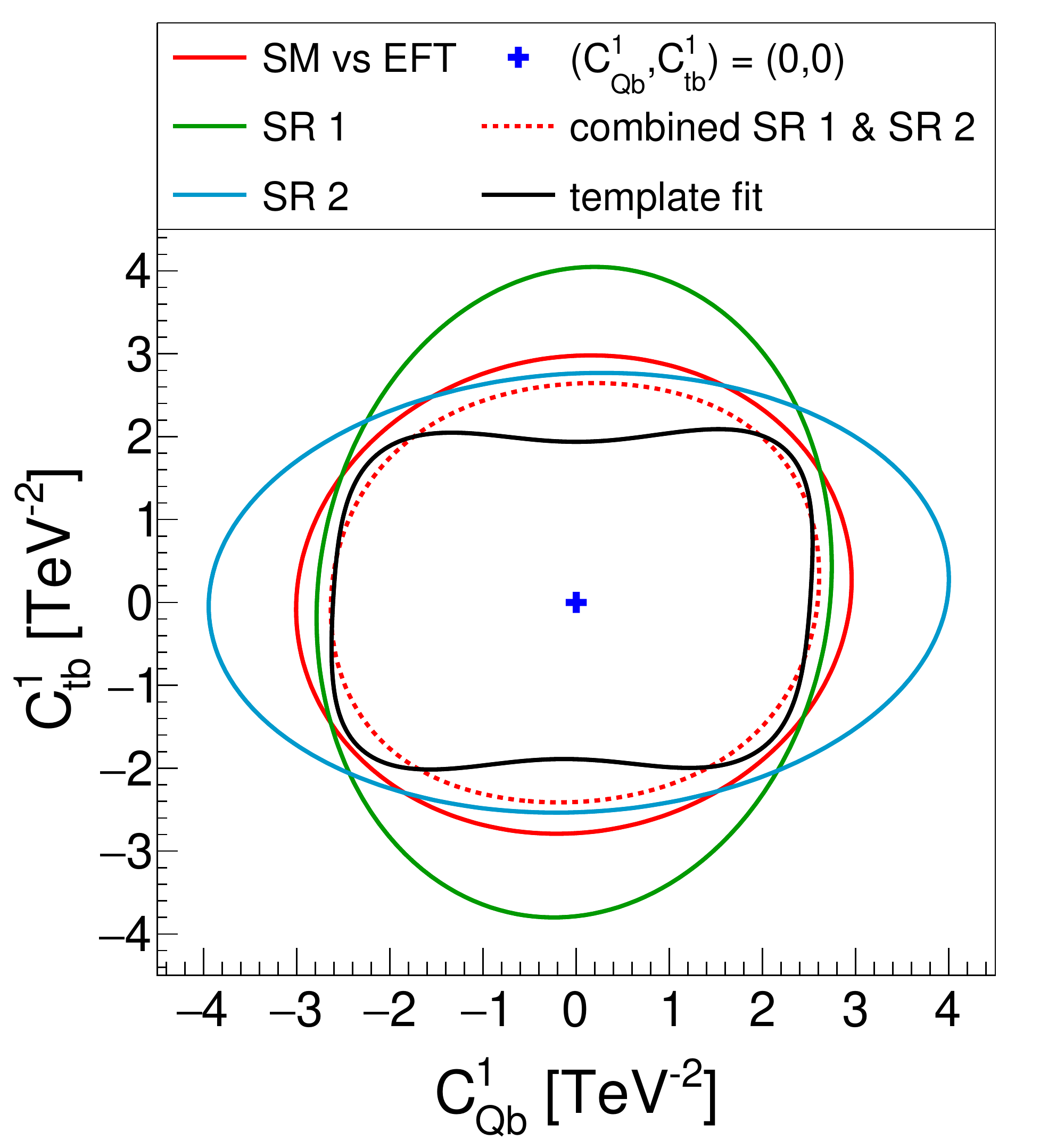}
\includegraphics[width=.4\textwidth]{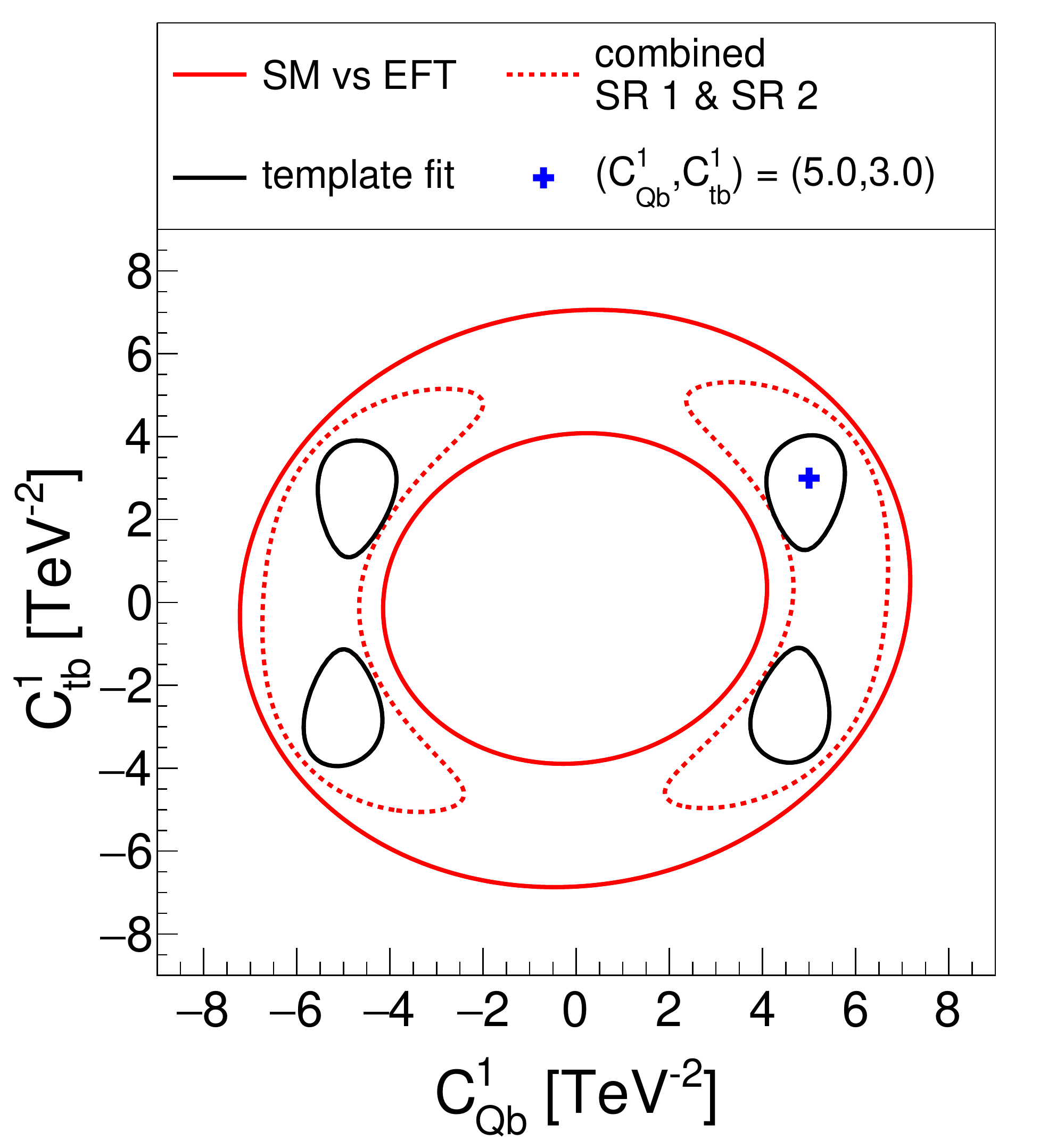}
\caption{\label{fig:Marginalized}
Limits at 95\% CL assuming a measurement consistent with the SM-only hypothesis (left) and the injection of a hypothestical EFT signal (right). 
}
\end{figure}

We consider two scenarios in which limits are derived on the two operators under consideration. A first scenario assumes an observation consistent with the SM prediction, such that limits are derived for the corresponding Wilson coefficients. In a second scenario, we inject a hypothetical signal corresponding to $C_{Qb}^{1} = 5$ TeV$^{-2}$ and $C_{tb}^{1} = 3$ TeV$^{-2}$ and we construct confidence intervals for the Wilson coefficients.
As a starting point, one can make a single selection on $P(t_{L}) + P(t_{R})$ asking this value to be larger than 0.83 (chosen to optimize the sensitivity), and perform the analysis in this EFT-enriched part of the phase space. The results are shown by the full red contours in Fig.~\ref{fig:Marginalized} on the left for the first scenario and on the right for the second scenario. It can clearly be seen that this selection is insensitive to the origin of the (hypothetical) excess, resulting in the toroidal shape of the two-dimensional confidence interval. We show that by deriving limits in the dedicated signal regions SR1 and SR2, and combining the obtained limits (shown by the dashed red lines), the sensitivity clearly improves. Finally, instead of selecting a part of the phase space based on the NN outputs, one can use the entire shape of the two-dimensional distributions in Fig.~\ref{fig:2DplotDiscriminators}. By performing binned maximum likelihood fits to predefined templates, the sensitivity is further improved and we are able to pinpoint the values of the Wilson coefficients with better precision (as shown by the black contours).

\section{conclusion}
We have presented a detailed investigation of extracting SMEFT signals in the $\ttbb$ final state. 
This process has shown itself to be a vital component  to constrain top EFT interactions.
Furthermore, our study serves as a proof of principle that motivates the use of multi-class discriminants in the context of globally constraining the SMEFT at the LHC. Such machine learning classifiers are capable of discriminating between operators mediating the production of top quarks in different helicity states.


\Acknowledgements
The authors would like to thank fruitful discussions with F. Maltoni, D. Pagani and G. Durieux. SM is an Aspirant van het Fonds Wetenschappelijk Onderzoek - Vlaanderen. CZ is supported by IHEP under Contract No. Y7515540U1. KM is supported by a Marie Sk\l{}odowska-Curie Individual Fellowship of the European Commission's Horizon 2020 Programme under contract number 707983.
AM is supported by the Strategic Research Program High-Energy Physics and the Research Council of the Vrije Universiteit Brussel, and by FWO under the ``Excellence of Science - EOS'' - be.h project n.30820817.

\end{document}